\documentstyle[twocolumn,aps,epsfig]{revtex}

\begin{document}

\draft

\title{Antiprotons at Solar Maximum
}
\author{John W. Bieber$^1$, R. A. Burger$^2$,
Ralph Engel$^1$, Thomas K. Gaisser$^1$,\\
Stefan Roesler$^3$, and Todor Stanev$^1$}
\address{
$^1$Bartol Research Institute, University of Delaware,
Newark, DE  19716, U.S.A.}
\address{
$^2$Space Research Unit, School of Physics, Potchefstroom University
for Christian Higher Education,\\
2520 Potchefstroom, South Africa}
\address{
$^3$CERN, CH-1211 Geneva 23, Switzerland}

\date{March 10, 1999}

\wideabs{
\maketitle

\begin{abstract}
\widetext
New measurements with good statistics will make it possible
to observe the time variation of cosmic antiprotons at 1 AU through
the approaching peak of solar activity.  We report a new computation
of the interstellar antiproton spectrum expected from collisions
between cosmic protons and the interstellar gas.  This spectrum
is then used as input to a steady-state drift model of solar modulation, 
in order to provide predictions for the antiproton spectrum as well as
the antiproton/proton ratio at 1 AU.  Our model predicts a
surprisingly large, rapid increase in the antiproton/proton
ratio through the next solar maximum, followed by a large
excursion in the ratio during the following decade.
\end{abstract}

\pacs{PACS numbers: 98.70.Sa, 96.40.Kk, 95.35.+d}

}

\narrowtext

\section{Introduction}
\label{sec:intro}

Two factors make cosmic antiprotons of special interest now.
First are new experimental results
\cite{IMAXpbar,CAPRICEpbar,MASSpbar,BESSpbar},
especially the abundant data from the 1997 flight of BESS \cite{Orito},
and the prospect of even more data from AMS \cite{Becker}
and future flights of BESS.
Second is the opportunity to observe with good statistics
the time variation of cosmic antiprotons at 1 AU through
a solar activity maximum.
These observations promise to provide new insights into
solar modulation, the process by which the expanding solar
wind modifies the energy spectrum of cosmic rays that
enter the heliosphere.  At the same time, any excess of
antiprotons that cannot be explained as a modulation effect
may point towards a primary source of antiprotons from
exotic processes \cite{MMO,Bottino98a,Bergstrom99a}.

Cosmic ray antiprotons are good probes of solar modulation
for several reasons.  First, the dominant process producing
antiprotons is collisions of high--energy
cosmic protons with the interstellar
gas.  The energy spectrum of antiprotons thus produced can be
computed with reasonable confidence, and we therefore have a
good {\it a priori} knowledge of the input spectrum for solar
modulation 
\cite{GaisserMaurer,Tan83a,GaisSchaef,Simon,Moskalenko98a,Bergstrom99a}.
Second, the antiproton production spectrum has a distinct
peak around 2 GeV kinetic energy because of the high energy
threshold for antiproton production in collisions.  This is
in sharp contrast \cite{Levy} to the featureless monotonic spectrum of
interstellar cosmic ray protons.  Third, because protons and
antiprotons differ only in charge sign, they are ideal for
studying solar modulation effects that depend explicitly
upon particle charge sign.

Existing evidence for charge sign dependent modulation appears
in Figure 1, which displays the ratio of cosmic electrons to
cosmic helium observed over a 25 year period
\cite{Garcia}, together with recent observations of the electron
to proton ratio made aboard Ulysses
\cite{Raviart}.
The largest variations are associated with reversals of
the Sun's magnetic polarity (shaded bars), which occur
near the peak of solar activity.
In 1970 and again in 1990, the charge ratios decreased rapidly.
In 1980 the ratio jumped upwards.  If the pattern continues,
another large, rapid increase in the negative/positive charge
ratio will occur through the polarity reversal expected
in 2000 or 2001.
\setcounter{figure}{0}
\begin{figure}[!hb]
\hspace*{-4mm}\epsfig{figure=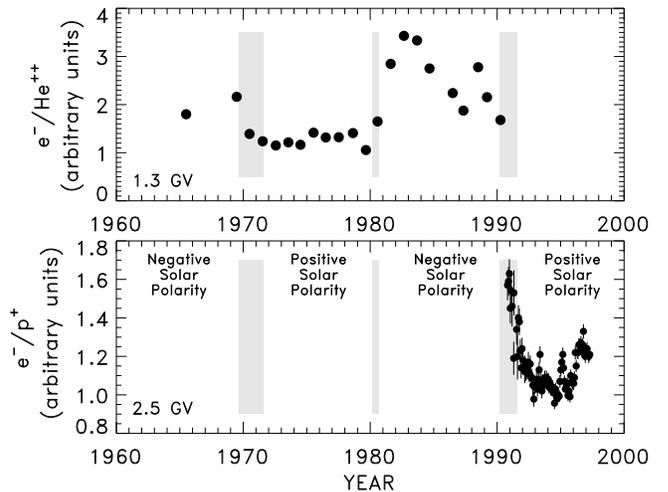,width=8.5cm}
\caption{Ratio of (top) cosmic electrons to cosmic helium at 1.3 GV rigidity
and (bottom) cosmic electrons to cosmic protons at 2.5 GV
rigidity.  Shaded areas delimit time periods when the Sun's
poloidal magnetic field was reversing.  Positive and negative
solar polarity refer to epochs when the magnetic field emerging
from the Sun's north pole point respectively outward and inward.
Data are from \protect\cite{Garcia,Raviart}.}
\label{fig1}
\end{figure}

In the remainder of this {\it Letter}, we present a new
calculation of the interstellar antiproton spectrum.
This spectrum is then used as input to a drift model
of solar modulation.  Finally,
we present our prediction for the evolution of the
antiproton spectrum and the antiproton/proton ratio
through the upcoming solar maximum.

\section{Interstellar antiproton spectrum}
\label{sec:pbar-prod}

In the framework of the standard leaky box model the continuity equation
describing secondary antiproton production \cite{GaisSchaef}
can be written as
\begin{equation}
\frac{1}{\lambda_e} J_{\bar p}(E_{\bar p})
+ \frac{1}{\lambda_i} J_{\bar p}(E_{\bar p})
= \frac{c}{4 \pi \langle m\rangle} Q(E_{\bar p}; J_{\bar p}(E_{\bar p})),
\end{equation}
where $\lambda_e$ is the characteristic escape length,
$J_{\bar p}(E_{\bar p})$ denotes the antiproton intensity, and
$\lambda_i$ is the interaction length for inelastic
collisions of antiprotons with the interstellar gas (annihilation
plus non-annihilation).
The mean free path length is
$ \lambda_i(E_{\bar p}) = \langle m\rangle/
\langle  \sigma^{\rm inel}_{\bar p}(E_{\bar p}) \rangle$,
where $\langle m\rangle$ and $\langle  \sigma^{\rm inel}_{\bar
p}(E_{\bar p}) \rangle$ denote the target mass and inelastic
cross section averaged over
the composition of the interstellar gas, respectively.
The mean escape length $\lambda_e$ is taken from the recent fit to
ratios of secondary to primary nuclei by Webber {\it et al.} \cite{Webber96}.

The source term $Q$ is split into two parts \cite{Simon}
$
Q(E_{\bar p}; J_{\bar p}(E_{\bar p})) =
Q_{\rm prod}(E_{\bar p}) + Q_{\rm scatt}(E_{\bar p})$.
Here, $Q_{\rm prod}$ is the source function for the production of
antiprotons due to collisions of primary cosmic rays with the
interstellar gas
\begin{equation}
Q_{\rm prod}(E_{\bar p}) = \frac{4 \pi}{c}\sum_{i,j} n_j \int_{E_{\rm
th}}^\infty
\frac{2\,d\sigma_{i,j\rightarrow\bar{p}}
}{dE_{\bar p}} J_i(E_i)dE_i\ ,
\end{equation}
and $Q_{\rm scatt}$ takes the inelastic scattering of antiprotons on the
interstellar gas into account
\begin{eqnarray}
Q_{\rm scatt}(E_{\bar p}) =
\nonumber\\
& & \hspace*{-2cm} 
\frac{4 \pi}{c}\sum_j n_j \int_{E_{\bar p}}^\infty 
\left\{ 
\frac{d\sigma_{\bar{p},j\rightarrow\bar{p}}}{dE_{\bar{p}}}
+
\frac{d\sigma_{\bar{p},j\rightarrow\bar{n}}}{dE_{\bar{n}}}
\right\} 
J_{\bar p}(E) dE\ .
\end{eqnarray}
The index $i$ sums over primary cosmic ray particles (protons and
alpha-particles in our calculation) and $j$ runs over all interstellar
gas target particle species (H, He, C, N, and O). The particle
abundances $n_j$ with $\sum_j n_j = 1$ are taken from the data compiled in
\cite{Meyer85a}. The antiproton production
and inelastic scattering cross sections have been calculated with a new
version of the {\sc Dtunuc} Monte Carlo event generator
\cite{Ferrari96a,Roesler98} which uses {\sc Phojet}
\cite{Engel95d} for the
simulation of elementary nucleon-nucleon collisions.

\section{Modulation code}

The effect of gradient and curvature drifts on solar modulation
has been intensively studied over the past 25 years
\cite{Jokipii,Kota,Moraal,Webber89}. Drifts in principle can provide
a natural explanation for charge sign dependent modulation effects
\cite{Potgieter90}, because particles with opposite charge
drift in opposite directions.
However, in recent years there has been an emerging consensus
that drifts may be important for modulation during low solar
activity, but that they become unimportant for several years
around solar maximum, owing to the disordered magnetic structure
of the heliosphere at that time \cite{Haasbroek,Potgieter}.

Recent work \cite{Burger} has challenged the
conventional wisdom that drifts can be ignored during
high solar activity.  This work finds that drifts produce
a strong differentiation between modulation of positive and
negative charges even during high solar activity.
There may be a brief interval during the polarity reversal
when the heliosphere is in a ``no drift'' state, but the
approach to and through this state is abrupt.
The observational evidence displayed in Figure 1 favors
this point of view decisively.  Indeed, the largest variation of
charge ratios occurs during peak solar activity in association
with the polarity reversal.

The principal factors governing solar modulation are
solar wind speed, tilt of the heliospheric current sheet,
and the cosmic ray diffusion tensor (which also embodies the drift
effect in its off--diagonal terms).
For wind speed, we use a simple latitude dependent model
consistent with the average properties of the solar wind \cite{Burger}.

The heliospheric current sheet is the surface in the solar
wind that separates opposing magnetic polarities.
It is essentially always tilted with respect to the solar
equator, by an angle that varies from about $10^\circ$
at sunspot minimum to more than $70^\circ$ during high
solar activity.  The combination of solar rotation and
radial expansion distorts the current sheet into a wavy
``ballerina skirt'' pattern,
and this complex field pattern is one of the factors
that leads to greater modulation during high solar
activity \cite{Thomas}.

The diffusion tensor{\bf \ }{\sf K }for a coordinate system with one axis
parallel to the background magnetic field, ${\bf B}_{o}$, and the other
two perpendicular to it, is
\begin{equation}
{\sf K}{\bf =}\left[
\begin{array}{lll}
\kappa _{||} & 0 & 0 \\
0 & \kappa _{\perp ,{\rm polar}} & \kappa _{T} \\
0 & -\kappa _{T} & \kappa _{\perp ,r\phi}
\end{array}
\right]
\end{equation}
with $\kappa _{||}$ the diffusion coefficient describing the diffusion
parallel to ${\bf B}_{o}$, $\kappa _{\perp ,{\rm polar}}$ and 
$\kappa_{\perp ,r\phi}$ the diffusion coefficients describing the
diffusion
perpendicular to the background magnetic field in the polar and
radial/azimuthal directions respectively, and $\kappa _{T}$ the 
drift coefficient. We use
quasi-linear theory and a slab/two-dimensional geometry for the
turbulence
to calculate $\kappa _{||}$ \cite{Bieber94a} and recent theoretical
results \cite{Zank96a} for the spatial variation of the fluctuations
in
the field as well as its correlation length, 
both quantities which appear in
$\kappa _{||}$. The perpendicular diffusion coefficients, $\kappa _{\perp ,%
{\rm polar}}$ and $\kappa _{\perp ,r\phi}$, are assumed to have
the same spatial dependence as $\kappa _{||}$, but a different
rigidity ($R$) dependence. 
For the drift coefficient $\kappa _{T{\rm }}$ we
assume that $\kappa _{T{\rm  }}/${\it particle speed} is proportional 
to $R^{3}$
at low rigidity, rolling over to $R^{1}$ at high rigidity. A detailed
discussion of a diffusion tensor similar to the one used in this work
can be found elsewhere \cite{Burger}.

Diffusion parameters were determined by fitting
model results to proton observations at 1 AU.
This procedure starts from an assumed interstellar proton spectrum.
Because a series of recent results \cite{LEAP,IMAX,CAPRICE,MASS,BESS}
indicates that the proton spectrum
in the range 10 to 100 GeV/nucleon is significantly lower than
previously assumed \cite{Webber}, we have correspondingly revised
downward the standard interstellar proton spectrum \cite{WebberIS}
in this energy range.  Our assumed interstellar hydrogen spectrum,
shown by the solid line in Fig. 2a, fits smoothly to the original
result of \cite{WebberIS} in the low energy region and to the data
\cite{LEAP,IMAX,CAPRICE,MASS,BESS} at high energy.

The fit to the solar minimum proton data \cite{Reinecke93a}
shown in Figure 2a, is obtained with a tilt angle of 10${{}^\circ}.$ To
model spectra near solar maximum, a tilt angle of 70${{}^\circ}$ is
used \cite{leRoux}
while $\kappa _{||},$ $\kappa_{\perp ,{\rm polar}}$ and 
$\kappa_{\perp ,r\phi}$ are all reduced by a factor of 2/3. 
Only the sign of the magnetic field is changed when going from one 
solar polarity epoch to the
other, i.e. it is only the direction in which particles drift that
distinguishes the two polarity states in the current study.

\section{Results}


In Fig. 2b the prediction for the local interstellar antiproton
intensity is shown.
Our calculation is in good agreement with the results reported in
\cite{Simon}, with a slightly higher value below one GeV, and it agrees
very well with the calculation of \cite{Bergstrom99a} at all
energies. We show here only the central value of our
calculation.  The theoretical uncertainties due to the limited
experimental data available for several quantities entering
the antiproton intensity
calculation will be discussed elsewhere \cite{Engel99ip},
as well the small differences with other calculations.
The antiproton spectra at 1AU are also shown in Fig. 2b with the
same coding as in Fig. 2a, and Fig. 2c shows the corresponding $\bar
p/p$ ratios.


Figure 3 displays the predicted dependence of the proton and
antiproton intensity at 1 AU (relative to interstellar level)
upon tilt angle of the heliospheric
current sheet, as well as the predicted dependence of their
ratio. 
The abscissa values have been arranged so that the
curves have the appearance of two successive solar cycles evolving in
time.  At the solar maximum of 2000, the solar
polarity switches from positive to negative.
The upper panel displays a well known
feature of drift models \cite{Kota}: the curves for positive charges are
broad during epochs of positive solar
polarity (1990's), and pointy during epochs of negative polarity
(decade beginning in 2000).  The opposite relationship holds
for negative charges.

Another difference is that protons
have a greater modulation amplitude ($\sim 4 \times$ between
solar maximum and solar minimum) than do antiprotons ($\sim 2 \times$)
\cite{Labrador97a}.
This stems from the differing character of their input spectra:  The
interstellar antiproton spectrum, with its peak around 2 GeV, is
``hard'' compared to the proton spectrum which has many
particles below 2 GeV.
This feature \cite{Levy} has been suggested
by \cite{MMO} as a way to distinguish a component of primary
antiprotons from an exotic source (such as annihilation of neutralinos
in the galactic halo or evaporation from primordial black holes) from
the galactic secondary component discussed in this paper.   Realization
of this idea depends on the extent to which the hypothetical
primary spectra are softer than the secondary spectra.  Recent
calculations of the spectra from neutralino annihilation
\cite{Bottino98a,Bergstrom99a} show
progressively harder spectra, making this signature correspondingly
less helpful.
\setcounter{figure}{2}
\begin{figure}
\centerline{\epsfig{figure=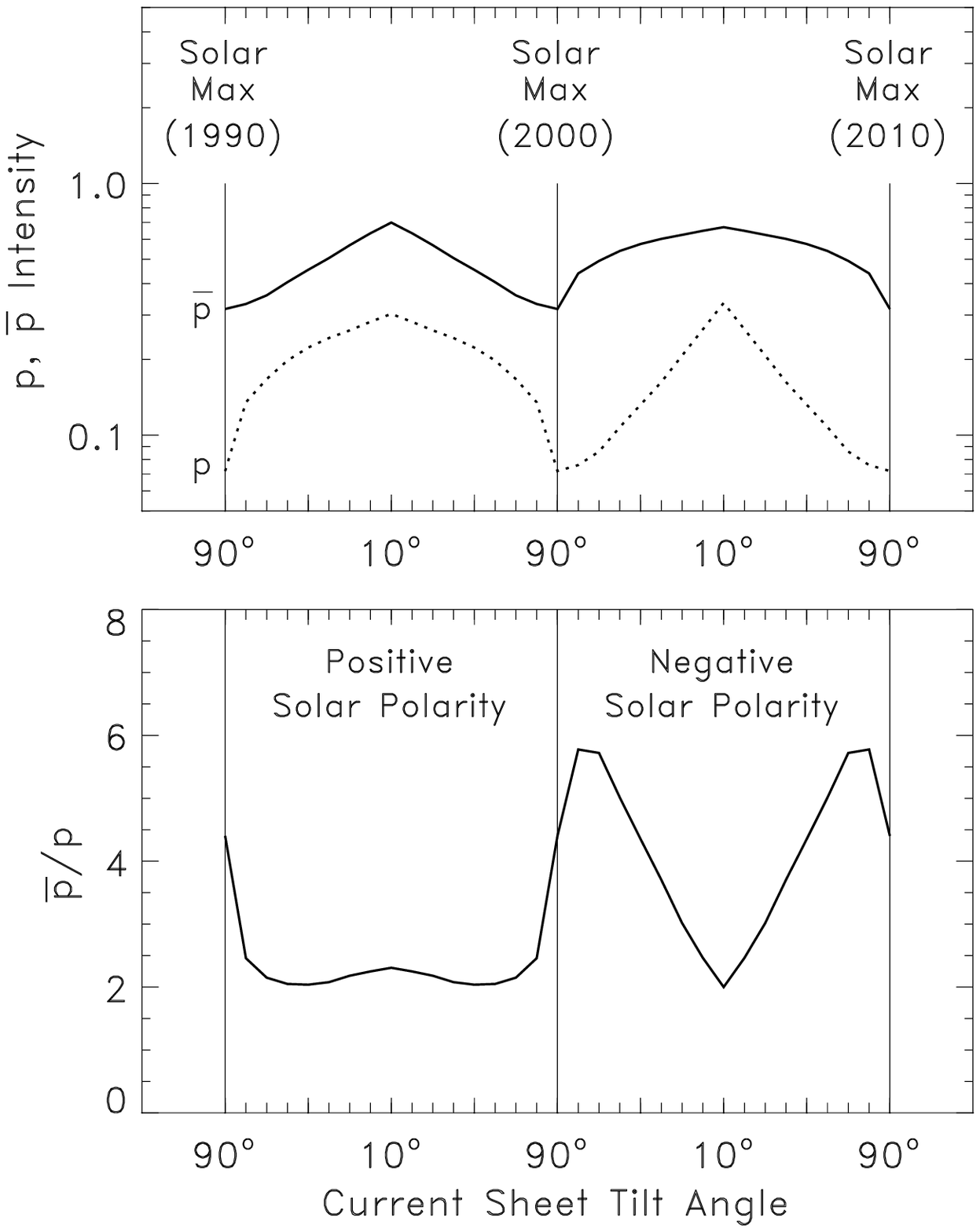,width=6.5cm,height=7cm}}
\caption{Predicted dependence of (top) the proton and antiproton intensity
and (bottom) the antiproton/proton ratio at 1 AU upon the tilt angle of the
heliospheric current sheet for 1 GeV kinetic energy. 
Intensities are relative to interstellar level.
Abscissa values are arranged
so that the curves mimic the expected time variation through two
solar cycles of opposite magnetic polarity (1990 to 2010).}
\label{fig3}
\end{figure}

\section{Summary}

Secondary galactic antiprotons provide a powerful probe of solar
modulation.   
Protons and antiprotons have sharply different interstellar spectra. 
They also drift in opposite directions because of their opposite charge
signs. The combination of these effects 
implies that the antiproton/proton ratio should display a much more
interesting evolution during the next 10 years than it did during the
1990's, when the ratio was nearly constant.
As we proceed through the sunspot maximum and
polarity switch expected about 2000, we predict that this ratio
will rapidly increase by a factor of about 3.  
Then, during the following decade,
it will display a large excursion closely tied to the variation
of the current sheet tilt angle. Actual
observation of these variations would be a stunning validation
of the importance of drift effects in solar
modulation at all phases of the solar activity cycle.


\noindent{\bf Acknowledgments:}
This work is supported by NASA Grant NAG5-5181. JWB is supported by NSF
grant ATM-9616610 and by NASA grant NAG5-7142. RAB and SR thank the
Bartol Res. Inst. for the hospitality.

\clearpage
\widetext

\setcounter{figure}{1}
\begin{figure}
\centerline{\epsfig{figure=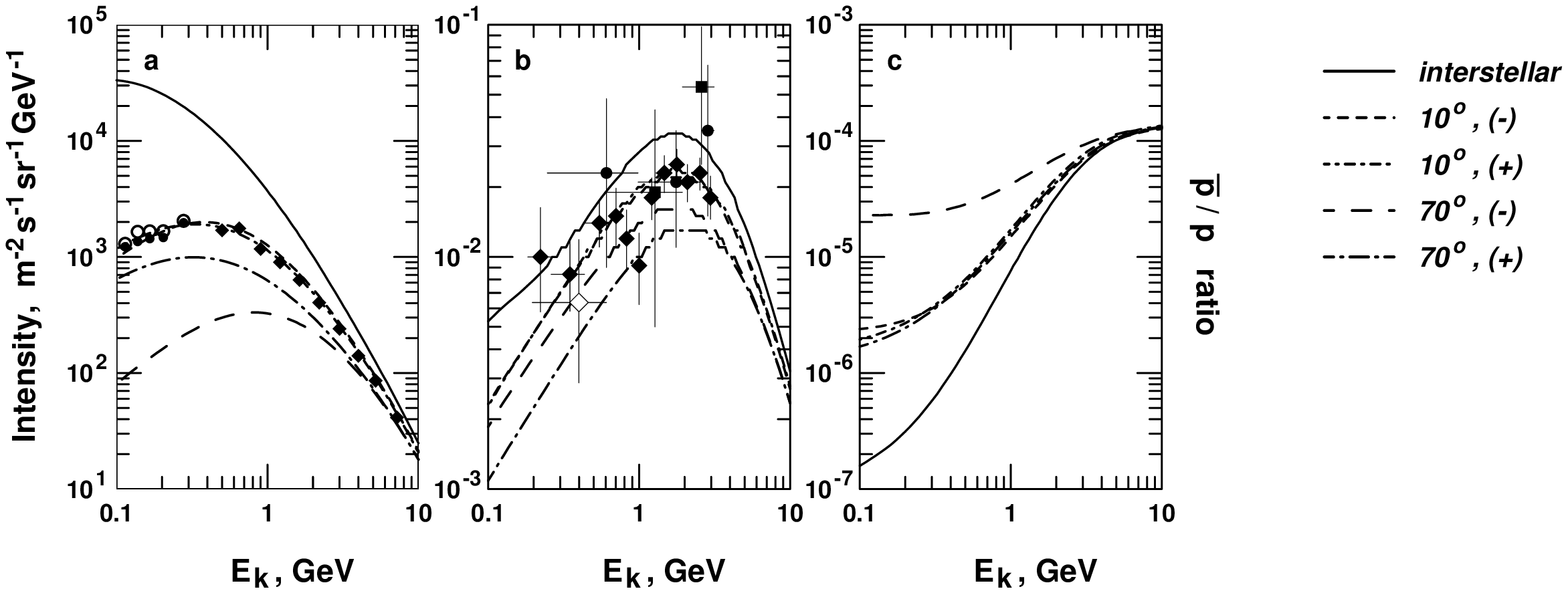,height=16cm,angle=90}}
\caption{Particle intensities: a) protons, b) antiprotons, and c)
ratio of antiproton to proton intensities. The data are taken from
\protect\cite{LEAP,IMAX,CAPRICE,MASS,BESS} for protons and
from \protect\cite{IMAXpbar,CAPRICEpbar,MASSpbar,BESSpbar} for
antiprotons.}
\label{fig2}
\end{figure}
\narrowtext

\end{document}